\documentclass[10pt]{article}

\usepackage{latexsym,axodraw,amssymb}

\newcommand{\eqalign}[1]
{\hspace{-10pt}\begin{array}{ll}#1\end{array}\hspace{-10pt}}

\def\bea{\begin{eqnarray}}
\def\eea{\end{eqnarray}}
\def\be{\begin{equation}}
\def\ee{\end{equation}}

\def\c{\hspace{-5pt}}
\def\Z{{\bf Z}}
\def\5{\bar 5}

\begin{document}
\pagestyle{empty}
\begin{center}
\hfill SISSA-74/2002/EP \\

\begin{center}
{\Large\bf Open string models with Scherk-Schwarz \\[2mm]
SUSY breaking}
\end{center}

\vspace{.2cm}
M. Trapletti

{\em ISAS-SISSA, Via Beirut 2-4, I-34013 Trieste, and INFN, Italy}
\end{center}

\centerline{\bf Abstract}
\vspace{-3 mm}
\begin{quote}\small
We apply the well-known Scherk-Schwarz supersymmetry breaking mechanism 
 in an open string context.
We construct a new $\Z_3\times \Z_3^\prime$ model,
containing only $D9$-branes, and rederive from a more geometric
perspective the known $\Z_6^\prime\times \Z_2^\prime$ model,
containing $D9$, $D5$ and $\bar D 5$ branes.
We show recent results about the study of quantum instability of these models.
\end{quote}

In the last years great efforts have been devoted to studying a way to embed the well-established 
knowledges about the Standard Model (SM) in a more fundamental microscopic theory.
String theory is one of the most promising candidates along this path, but we do not
have a complete solution to the problem yet.

In this pattern supersymmetry (SUSY) plays a crucial role, 
for example explaining how the the hierarchy problem can be solved, but also stabilizing the various
string models one can build.
It is also clear that a phenomenologically appealing string model must contain a mechanism that breaks SUSY at
a suitable scale (TeV), and this make things difficult, because it is extremely hard to build 
truly stable non-SUSY vacua in string theory.

To this purpose we have taken into account  
the so-called Scherk-Schwarz (SS) symmetry-breaking 
mechanism \cite{SS}, applicable on theories with compact extra dimensions.
As described in the next section, it consists in suitably twisting the periodicity conditions
of each field along some compact directions. In this way, one obtains a non-local, perturbative
and calculable symmetry breaking mechanism. String models of this type can be constructed by
deforming supersymmetric orbifold models \cite{orb}; a variety of four-dimensional (4D)
closed string models, mainly based on $\Z_2$ orbifolds, have been constructed in this way
\cite{ssstringa}. More in general, SS symmetry breaking can be achieved through freely-acting
orbifold projections \cite{kk}. This fact has been recently exploited in \cite{root} to construct
a novel class of closed string examples, including a model based on a $\Z_3$ orbifold.
Unfortunately, a low compactification scale is quite unnatural for closed string models, where
the fundamental string scale $M_s$ is tied to the Planck one, and can be achieved only in very
specific situations \cite{large} (see also \cite{amq}). The situation is different for open strings,
where $M_s$ can be very low \cite{mill}, and interesting open string models with SS SUSY breaking
have been derived in \cite{ads1,adds,cotrone}. Recently, the SS mechanism has been the object of renewed 
interest also from a more phenomenological ``bottom-up'' viewpoint, where it has been used in 
combination with orbifold projections to construct realistic 5D non-SUSY extensions of the SM 
\cite{pom,bhn}.

We will describe the general ideas proposed in \cite{root} and build 
chiral IIB orientifold models with SS supersymmetry breaking.
The most appealing common feature of these models is that they are tachyon-free for a suitable choice of the
compactification moduli, so that instabilities, at least at the classical level, are still
avoided. Recently
new studies \cite{nuovo} have been performed to analyze the quantum stability more deeply, in particular the potential
for the crucial moduli, which is flat at tree level, have been computed at one loop, showing a good behavior, at
least for the model based on the  $\Z_2$ orbifold.

\section{The Scherk-Schwarz mechanism in string theory}
The Scherk-Schwarz mechanism was introduced in models where some compact extra dimension is present.
Given a symmetric theory  under a group $G$ it is possible to break the symmetry by fixing different boundary
conditions for fields in the same multiplet.
If we consider a SUSY theory with fields $\phi_F$ defined on the compact dimension 
$x\sim x+2\pi R$, where $F$ labels the fermionic or bosonic nature of each field, the procedure of 
SUSY breaking is encoded in the twisted boundary conditions
\bea
\phi_F(x+2\pi R)=g(F)\phi_F(x),
\eea
where $g(F)$ is, for example, a phase that takes different values for bosons and fermions.
In a compact formalism, introducing $P$ as a translation along the compact dimension,
\bea
\phi_F=g(F)P\phi_F.
\eea
In string theory this is realized through an orbifold projection with an operator obtained by
joining a SUSY-breaking operator $g$ with a translation along a compact dimension.

The key feature is that the translation shifts the tachyon masses that usually arise when
orbifolding with $g$ only. The shift is proportional to $R^2$, $R$ being the radius of the compact direction
where the translation acts, so that, for $R$ sufficiently big the model is tachyon free.
Moreover the scale of SUSY  breaking is proportional to $R$.

We demand that $g$ and $P$ commute, and so we also need the two operators to have 
order $N_g$ and $N_P$ such that $n N_g=N_P$ for some $n\in\mathbb{N}$. 
This is due to the fact that modular invariance requires, in the closed sector,
the introduction of sector twisted by $(gP)^q$, with $q\in\{0,1,\dots,N_{gP}-1\}$, 
so that if the above relation is not fulfilled there exists at least one 
sector twisted only by $g$ that contains tachyons.
In the next sections we shall introduce an order-two and an order-three operator and  apply 
them to $d=4$, $N=1$ string models.

\section{\label{Z6} The $\Z_6^\prime \times \Z_2^\prime$ model}

The $\Z_6^\prime \times \Z_2^\prime$ orientifold of \cite{adds} is
obtained by applying a SUSY-breaking $\Z_2^\prime$ projection to the
SUSY $\Z_6^\prime$ model of \cite{afiv}. The $\Z_6^\prime$ group is generated
by $\theta$ acting as rotations of angles $2\pi v^\theta_i$ in the  three
internal tori $T^2_i$ ($i=1,2,3$), with $v^\theta_i=1/6 (1,-3,2)$. The
$\Z_2^\prime$ group is instead generated by $\beta$ acting as a translation
of length $\pi R$ along one of the radii of $T^2_2$ (that we shall call SS
direction in the following), combined with a sign $(-)^F$, where $F$ is the
4D space-time fermion number. Beside the $O9$-plane, the model contains $O5$-planes
at $y=0$ and $y=\pi R$ (as the corresponding
SUSY model \cite{afiv}) and $\bar O5$-planes
at $y=\pi R/2$ and $y=3\pi R/2$ along the SS direction (see Fig.~1),
corresponding to the two order-2 elements $\theta^3$ and $\theta^3\beta$.
In order to cancel both NSNS and RR massless tadpoles, $D9$, $D5$ and $\bar D 5$-branes
must be introduced.
\begin{figure}[h]
\begin{center}
\begin{picture}(280,80)(0,0)
\Line(0,0)(75,0)
\Line(37.5,65)(112.5,65)
\Line(0,0)(37.5,65)
\Line(75,0)(112.5,65)
\Vertex(0,0){2}
\put(8,5){\footnotesize ${{}_{1=1^\prime}}$}
\Vertex(37.5,21.65){2}
\put(45.5,26.65){\footnotesize${ {}_2}$}
\Vertex(75,43.3){2}
\put(83, 48.3){\footnotesize${ {}_3}$}
\Vertex(37.5,0){2}
\put(45,5){\footnotesize${ {}_{2^\prime}}$}
\Vertex(56.25,32.48){2}
\put(64.25,37.48){\footnotesize${ {}_{3^\prime}}$}
\Vertex(18.75,32.48){2}
\put(26.75,37.48){\footnotesize${ {}_{4^\prime}}$}
\Line(200,0)(275,0)
\Line(237.5,65)(312.5,65)
\Line(200,0)(237.5,65)
\Line(275,0)(312.5,65)
\Vertex(200,0){2}
\put(208,5){\footnotesize${ {}_1}$}
\Vertex(237.5,21.65){2}
\put(245.5,26.65){\footnotesize${ {}_2}$}
\Vertex(275,43.3){2}
\put(283, 48.3){\footnotesize${ {}_3}$}
\Line(100,0)(175,0)
\Line(137.5,65)(212.5,65)
\Line(100,0)(137.5,65)
\Line(175,0)(212.5,65)
\Vertex(100,0){2}
\put(108,5){\footnotesize${ {}_1}$}
\Vertex(137.5,0){2}
\put(145,5){\footnotesize${ {}_2}$}
\Vertex(156.25,32.48){2}
\put(164.25,37.48){\footnotesize${ {}_3}$}
\Vertex(118.75,32.48){2}
\put(126.75,37.48){\footnotesize${ {}_4}$}
\Vertex(118.75,0){2}
\put(126.75,5){\footnotesize${ {}_{1^\prime}}$}
\Vertex(156.25,0){2}
\put(164.25,5){\footnotesize${ {}_{2^\prime}}$}
\Vertex(175,32.48){2}
\put(183,37.48){\footnotesize${ {}_{3^\prime}}$}
\Vertex(137.5,32.48){2}
\put(145.5,37.48){\footnotesize${ {}_{4^\prime}}$}
\end{picture}
\caption{\label{fix1}
{\footnotesize \em  The fixed-points structure in the $\Z_6^\prime \times \Z_2^\prime$ model.
We label the 12 $\theta$-fixed points with $P_{1bc}$ and the 12 $\theta\beta$-fixed
points with $P_{1bc^\prime}$, each index referring to a $T^2$, ordered as in the figure.
Similarly, we denote with
$P_{a\bullet c}$ the 9 $\theta^2$-fixed planes filling the second $T^2$, and
respectively with
$P_{a^\prime b\bullet }$ and $P_{a^\prime  b^\prime \bullet}$ the 16 $\theta^3$-fixed
and $\theta^3\beta$-fixed planes filling the third $T^2$. The 32 $D5$-branes
and the 32 $\bar D 5$-branes are located at the point 1 in the first $T^2$, fill the
third $T^2$, and sit at the points $1$ and $1^\prime$ respectively in the second $T^2$.}}
\vskip -10pt
\end{center}
\end{figure}
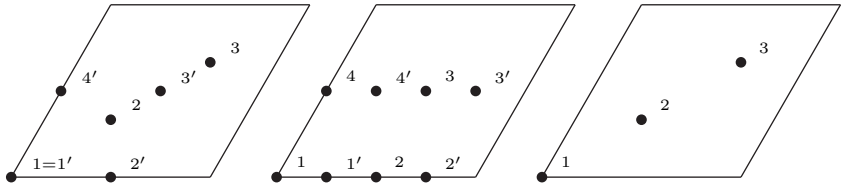

\subsection{Spectrum}
The main features of the closed string spectrum of the $\Z_6^\prime \times \Z_2^\prime$
model can be deduced from those of the $\Z_6^\prime$ model, that can be found in \cite{afiv}.
The spectrum is summarized in
Table 1.
\begin{table}[t]
\vbox{
$$\vbox{\offinterlineskip
\hrule height 1.1pt
\halign{&\vrule width 1.1pt#
&\strut\quad#\hfil\quad&
\vrule width 1.1pt#
&\strut\quad#\hfil\quad&
\vrule#
&\strut\quad#\hfil\quad&
\vrule width 1.1pt#\cr
height3pt
&\omit&
&\omit&
&\omit&
\cr
&\hfil ${\rm Sector}$ &
&\hfil  $\c\Z_6^\prime \times \Z_2^\prime$&
&\hfil  $\c\Z_3 \times \Z_3^\prime$&
\cr
height3pt
&\omit&
&\omit&
&\omit&
\cr
\noalign{\hrule
}
height3pt
&\omit&
&\omit&
&\omit&
\cr
& \hfil Untwisted&
&\hfil 1 graviton, 5 scalars&
&\hfil 1 graviton, 11 scalars, 1+1 spinors&
\cr
height3pt
&\omit&
&\omit&
&\omit&
\cr
\noalign{\hrule}
height3pt
&\omit&
&\omit&
&\omit&
\cr
&\hfil $\theta$ twisted&
&\hfil 6 chiral multiplets &
&\hfil 6 hypermultiplets &
\cr
height3pt
&\omit&
&\omit&
&\omit&
\cr
\noalign{\hrule}
height3pt
&\omit&
&\omit&
&\omit&
\cr
&\hfil $\theta^2$ twisted&
&\hfil 18 scalars&
&\hfil ${\bf -}$ &
\cr
height3pt
&\omit&
&\omit&
&\omit&
\cr
\noalign{\hrule}
height3pt
&\omit&
&\omit&
&\omit&
\cr
&\hfil $\theta^3$ twisted&
&\hfil 6 chiral multiplets  &
&\hfil ${\bf -}$&
\cr
height3pt
&\omit&
&\omit&
&\omit&
\cr
\noalign{\hrule}
height3pt
&\omit&
&\omit&
&\omit&
\cr
&\hfil $\theta\beta$ twisted&
&\hfil 6 chiral &
&\hfil 9 chiral &
\cr
height3pt
&\omit&
&\omit&
&\omit&
\cr
\noalign{\hrule}
height3pt
&\omit&
&\omit&
&\omit&
\cr
&\hfil $\theta^3\beta$ twisted&
&\hfil 6 chiral &
&\hfil ${\bf -}$ &
\cr
height3pt
&\omit&
&\omit&
&\omit&
\cr
\noalign{\hrule}
height3pt
&\omit&
&\omit&
&\omit&
\cr
&\hfil $\theta\beta^2$ twisted&
&\hfil \hfil ${\bf -}$ &
&\hfil 9 chiral &
\cr
height3pt
&\omit&
&\omit&
&\omit&
\cr
}
\hrule height 1.1pt}
$$
}
\caption{{\footnotesize \em Massless closed string spectrum for $\Z_6^\prime \times \Z_2^\prime$
and $\Z_3 \times \Z_3^\prime$  models. We used $\theta$ as the generator
of $\Z_6^\prime$ ($\Z_3$) and $\beta$ as the generator of $\Z_2^\prime$ ($\Z_3^\prime$).
Hypermultiplets are multiplets of N=1 SUSY in 6D, while chiral
multiplets are multiplets of N=1 SUSY in 4D.
The SUSY generators are different in the different sectors as explained
in the text.
The two spinors in the untwisted sector of $\Z_3 \times \Z_3^\prime$ have opposite
chirality.}}
\label{closedspectrum}
\end{table}

The open spectrum sectors are defined by tadpole cancellation, that
requires the presence of $D9$, $D5$ and $\bar D 5$ branes.
Tadpole cancellation leaves a certain freedom in the choice of the gauge group, 
for simplicity we consider the case of maximal unbroken gauge symmetry where
all $D5$ and $\bar D 5$ branes are located respectively at $P_{11\bullet}$ and $P_{11^\prime\bullet}$
(see Fig.~1 and its caption). The $\Z_2^\prime$ projection requires then that an equal number of
image branes are located respectively at $P_{12\bullet}$ and $P_{12^\prime\bullet}$.
We do not consider the case in which branes and anti-branes coincide also along the SS direction
because this configuration is unstable even classically, due to the presence of open string tachyons.
On the other hand, fixing the branes at antipodal points along the SS direction allows a metastable
configuration without open string tachyons for sufficiently large SS radius.
The massless open string spectrum can now be easily derived, and is summarized in Table \ref{spectrum}.

\section{A $\Z_3 \times \Z_3^\prime$ model}

It has been shown in \cite{root} that SS symmetry breaking can be obtained
also in $\Z_3$ models through a suitable freely-acting and SUSY-breaking
$\Z_3^\prime$ projection. In this section, we will construct a new
$\Z_3 \times \Z_3^\prime$ model, based on this structure, that will prove
to be much simpler than the $\Z_6^\prime \times \Z_2^\prime$ one.

The $\Z_3\times\Z^\prime_3$ orbifold group is defined in the following way \cite{root}.
The $\Z_3$ generator $\alpha$ acts as a SUSY-preserving rotation with twist $v^\alpha_i=1/3(1,1,0)$,
while the $\Z_3^\prime$ generator $\beta$ acts as a SUSY-breaking rotation with
$v^\beta_i=1/3(0,0,2)$ and an order-three diagonal translation $\delta$ in $T^2_1$.

\begin{figure}[h!]
\begin{center}
\begin{picture}(280,85)(0,0)
\Line(200,0)(275,0)
\Line(237.5,65)(312.5,65)
\Line(200,0)(237.5,65)
\Line(275,0)(312.5,65)
\Vertex(200,0){2}
\put(208,5){\footnotesize${ {}_1}$}
\Vertex(237.5,21.65){2}
\put(245.5,26.65){\footnotesize${ {}_2}$}
\Vertex(275,43.3){2}
\put(283, 48.3){\footnotesize${ {}_3}$}
\Line(100,0)(175,0)
\Line(137.5,65)(212.5,65)
\Line(100,0)(137.5,65)
\Line(175,0)(212.5,65)
\Vertex(100,0){2}
\put(108,5){\footnotesize${ {}_1}$}
\Vertex(137.5,21.65){2}
\put(145.5,26.65){\footnotesize${ {}_2}$}
\Vertex(175,43.3){2}
\put(183,48.3){\footnotesize${ {}_3}$}
\Line(0,0)(75,0)
\Line(37.5,65)(112.5,65)
\Line(0,0)(37.5,65)
\Line(75,0)(112.5,65)
\Vertex(0,0){2}
\put(8,5){\footnotesize${ {}_1}$}
\Vertex(37.5,21.65){2}
\put(45.5,26.65){\footnotesize${ {}_2}$}
\Vertex(75,43.3){2}
\put(83,48.3){\footnotesize${ {}_3}$}
\Vertex(12.5,21.65){2}
\put(20.5,26.65){\footnotesize${ {}_{1^\prime}}$}
\Vertex(50,43.3){2}
\put(58,48.3){\footnotesize${ {}_{2^\prime}}$}
\Vertex(50,0){2}
\put(58,5){\footnotesize${ {}_{3^\prime}}$}
\Vertex(25,0){2}
\put(33,5){\footnotesize${ {}_{1^{\prime\prime}}}$}
\Vertex(62.5,21.65){2}
\put(70.5,26.65){\footnotesize${ {}_{2^{\prime\prime}}}$}
\Vertex(25,43.3){2}
\put(33,48.3){\footnotesize${ {}_{3^{\prime\prime}}}$}
\end{picture}
\caption{{\footnotesize \em The fixed-point structure in the $\Z_3 \times \Z_3^\prime$ model.
We label the 9 $\alpha$-fixed planes with $P_{ab\bullet}$, the
27 $\alpha\beta$-fixed points with $P_{a^\prime bc}$, the 27
$\alpha\beta^2$-fixed points with $P_{a^{\prime\prime} bc}$, and
the 3 $\beta$-fixed planes with $P_{\bullet\bullet c}$.}}
\vskip -10pt
\end{center}
\end{figure}
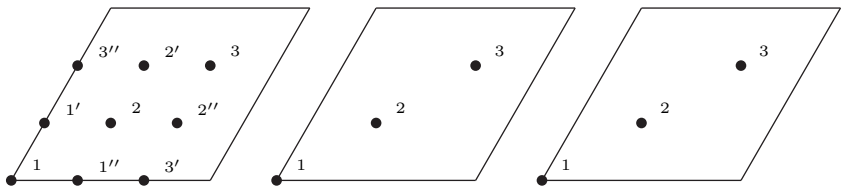

\subsection{Spectrum}
The closed string spectrum is summarized in Table \ref{closedspectrum}.
The open spectrum contains only a sector from strings stretching between $D9$-branes.
The maximal gauge group is $SO(8)\times U(8)\times U(4)$. The $U(8)\times U(4)$
factor comes from the $U(12)$ gauge factor of the 4D N=1 $\Z_3$ model, which is further broken by
the $\Z_3^\prime$ projection. As in the previous model, this can be interpreted as a Wilson
line symmetry breaking. 
Notice that all the gauginos have a mass $\sim 1/R_1$.
The spectrum of charged massless states is easily obtained and reported in Table \ref{spectrum}.

\begin{table}[h]
\vbox{
$$\vbox{\offinterlineskip
\hrule height 1.1pt
\halign{&\vrule width 1.1pt#
&\strut\quad#\hfil\quad&
\vrule width 1.1pt#
&\strut\quad#\hfil\quad&
\vrule#
&\strut\quad#\hfil\quad&
\vrule width 1.1pt#\cr
height3pt
&\omit&
&\omit&
&\omit&
\cr
&\hfil &
&\hfil  $\c\Z_6^\prime \times \Z_2^\prime$&
&\hfil  $\c\Z_3 \times \Z_3^\prime$&
\cr
height3pt
&\omit&
&\omit&
&\omit&
\cr
\noalign{\hrule
}
height3pt
&\omit&
&\omit&
&\omit&
\cr
&\hfil $G_9:$&
&\hfil$\eqalign{U(4)^2 \times U(8)}$&
&\hfil$\eqalign{SO(8) \times U(8)\times U(4)}$&
\cr
&\hfil $G_5=G_{\5}:$&
&\hfil$\eqalign{    U(2)^2 \times U(4)}$&
&\hfil ${\bf -}$&

\cr
height3pt
&\omit&
&\omit&
&\omit&
\cr
\noalign{\hrule}
height3pt
&\omit&
&\omit&
&\omit&
\cr
&\hfil 99 scalars&
&$\eqalign{
 {\bf (4,4,1),\;(\overline 4,\overline 4,1),\;(1,1,28),}\cr
 {\bf (1,1,\overline{28}),\;(6,1,1),\;(1,4,\overline 8),}\cr
 {\bf (\overline 4,1,8),\; (1,\overline 6,1),\;(\overline 4,4,1),}\cr
 {\bf (4,1,8),\;(1,\overline 4,\overline 8)}}$&
&$\eqalign{
2{\bf (8,8,1),\;}2{\bf (1,\overline {28},1),}\cr
{\bf (8,1,4),\;(1,1,\overline 6)}
}$&
\cr
height3pt
&\omit&
&\omit&
&\omit&
\cr
\noalign{\hrule}
height3pt
&\omit&
&\omit&
&\omit&
\cr
&\hfil 99\,\, fermions&
&\hfil ${\bf -}$&
& $\eqalign{
2{\bf (8,1,4),\;} 2{\bf (1,1,\overline 6),} \cr
{\bf (1,\overline {8},4),\;}
{\bf (1,\overline 8,\overline4)\;}\cr
}$   &
\cr
height3pt
&\omit&
&\omit&
&\omit&
\cr
\noalign{\hrule}
height3pt
&\omit&
&\omit&
&\omit&
\cr
&\hfil $55$ chiral mult.&
&$\eqalign{
 {\bf (2,2,1),\;(\overline 2,\overline 2,1),\;(1,1,6),}\cr
 {\bf (1,1,\overline 6),\;(1_A,1,1),\;(1,2,\overline 4),}\cr
 {\bf (\overline 2,1,4),\; (1,\overline 1_A,1),\;(\overline 2,2,1),}\cr
 {\bf (2,1,4),\;(1,\overline 2,\overline 4)}}$&
&\hfil ${\bf -}$&
\cr
height3pt
&\omit&
&\omit&
&\omit&
\cr
\noalign{\hrule}
height3pt
&\omit&
&\omit&
&\omit&
\cr
&\hfil $95$ chiral mult.&
&$\eqalign{
 {\bf (\overline 4,1,1;\overline 2,1,1),\;(1,4,1;1,2,1),}\cr
 {\bf (4,1,1;1,1,\overline 4),\;(1,1,\overline 8;2,1,1),}\cr
 {\bf (1,\overline 4,1;1,1,4),\;(1,1,8;1, \overline 2 ,1)}}$&
&\hfil ${\bf -}$&
\cr
height3pt
&\omit&
&\omit&
&\omit&
\cr
}
\hrule height 1.1pt}
$$
}
\caption{{\footnotesize \em Massless open string spectrum for $\Z_6^\prime \times \Z_2^\prime$
and $\Z_3 \times \Z_3^\prime$  models. In the 55 sector chiral multiplets
in the representation of $G_5$ are reported.
The matter content of the $\5\5$ sector is the same of the $55$
sector but in conjugate representations of $G_{\5}=G_5$.
In the 95 sectors chiral multiplets are present in representations of
$G_9\times G_5$ respectively. Again, the matter content in the $9\5$ sector
is obtained from that in the 95 sector by conjugation.}}
\label{spectrum}
\end{table}

\section{Discussion}
The Scherk-Schwarz SUSY breaking mechanism has been introduced in open string 
theory and applied to two $d=4$, $N=1$
models. We obtained two non-SUSY tachyon-free chiral models, whose 
stability has been discussed in \cite{nuovo}, where the potential $V$ for the 
relevant modulus $R$ defined in the first section has been studied. 
$V$ is flat at tree level but gets corrections from 
loop amplitudes. We considered the one-loop amplitudes for various models, in the odd-order orbifolds 
a quantum instability is found in every case, while in the  $Z_2$ case  $V$ shows a minimum. This 
stability has also been studied considering more than a modulus, and it seems that the presence of a 
general minimum is not guaranteed, but in this case we think that a deeper analysis is needed.

In \cite{paper} we have also studied the local anomaly cancellation in the two models.
All pure gauge and mixed gauge-gravitational anomalies cancel, thanks to a generalized
GS mechanism that involves also twisted RR 4-forms, necessary to cancel localized
irreducible 6-form terms in the anomaly polynomial, which vanish only globally.
The 4D remnant of this mechanism is a local Chern--Simons term. The local (and global)
cancellation of reducible anomalies is instead ensured by twisted RR axions. In the
latter case, even $U(1)$ gauge fields affected by anomalies that vanish only globally in 4D
are spontaneously broken by the GS mechanism. Such $U(1)$'s do not appear in the corresponding
SUSY $\Z_6^\prime$ and $\Z_3$ 4D orientifolds.

\vskip 25pt
\noindent
{\Large \bf Acknowledgments}
\vskip 10pt

\noindent
I would like to thank C.~A.~Scrucca, M.~Serone and M.~Borunda for the fruitful collaboration this work is
based on. I would like also to thank Daniele Perini for useful discussions.
This work was partially supported by the EC through the RTN network ``The quantum structure 
of space-time and the  geometric nature of fundamental interactions'', contract HPRN-CT-2000-00131.


\begin{thebibliography}{99}

\bibitem{SS}
J.~Scherk and J.~H.~Schwarz,
%``Spontaneous Breaking Of Supersymmetry Through Dimensional Reduction,''
Phys.\ Lett.\ B {\bf 82} (1979) 60;
%%CITATION = PHLTA,B82,60;%%
%``How To Get Masses From Extra Dimensions,''
Nucl.\ Phys.\ B {\bf 153} (1979) 61.
%%CITATION = NUPHA,B153,61;%%
%
\bibitem{orb}
L.~J.~Dixon, J.~A.~Harvey, C.~Vafa and E.~Witten,
%``Strings On Orbifolds,''
Nucl.\ Phys.\ B {\bf 261} (1985) 678;
%%CITATION = NUPHA,B261,678;%%
%``Strings On Orbifolds. 2,''
Nucl.\ Phys.\ B {\bf 274} (1986) 285.
%%CITATION = NUPHA,B274,285;%%
%
\bibitem{ssstringa}
C.~Kounnas and M.~Porrati,
%``Spontaneous Supersymmetry Breaking In String Theory,''
Nucl.\ Phys.\ B {\bf 310} (1988) 355; \\
%%CITATION = NUPHA,B310,355;%%
S.~Ferrara, C.~Kounnas, M.~Porrati and F.~Zwirner,
%``Superstrings With Spontaneously Broken Supersymmetry And Their Effective Theories,''
Nucl.Phys.B{\bf 318} (1989) 75; 
%%CITATION = NUPHA,B318,75;%%
%
\bibitem{kk}
R.~Rohm,
%``Spontaneous Supersymmetry Breaking In Supersymmetric String Theories,''
Nucl.\ Phys.\ B {\bf 237} (1984) 553; \\
%%CITATION = NUPHA,B237,553;%%
C.~Kounnas and B.~Rostand,
%``Coordinate Dependent Compactifications And Discrete Symmetries,''
Nucl.\ Phys.\ B {\bf 341} (1990) 641; \\
%%CITATION = NUPHA,B341,641;%%
E.~Kiritsis and C.~Kounnas,
%``Perturbative and non-perturbative partial supersymmetry breaking:  N = 4 $\to$ N = 2 $\to$ N = 1,''
Nucl.\ Phys.\ B {\bf 503} (1997) 117
[hep-th/9703059].
%%CITATION = HEP-TH 9703059;%%
%
\bibitem{root}
C.~A.~Scrucca and M.~Serone,
% ``A novel class of string models with Scherk-Schwarz supersymmetry breaking''
JHEP\ {\bf 0110} (2001) 017
[hep-th/0107159].
%
%\cite{Borunda:2002ra}
\bibitem{nuovo}
M.~Borunda, M.~Serone and M.~Trapletti,
%``On the quantum stability of type IIB orbifolds and orientifolds with  Scherk-Schwarz SUSY breaking,''
arXiv:hep-th/0210075.
%%CITATION = HEP-TH 0210075;%%
%
\bibitem{large}
I.~Antoniadis,
%``A Possible New Dimension At A Few Tev,''
Phys.\ Lett.\ B {\bf 246} (1990) 377.
%%CITATION = PHLTA,B246,377;%%
%
\bibitem{amq}
I.~Antoniadis, C.~Munoz and M.~Quiros,
%``Dynamical supersymmetry breaking with a large internal dimension,''
Nucl.\ Phys.\ B {\bf 397} (1993) 515
[hep-ph/9211309]; \\
%%CITATION = HEP-PH 9211309;%%
I.~Antoniadis and K.~Benakli,
%``Limits on extra dimensions in orbifold compactifications of superstrings,''
Phys.\ Lett.\ B {\bf 326} (1994) 69
[hep-th/9310151].
%%CITATION = HEP-TH 9310151;%%
%
\bibitem{mill}
I.~Antoniadis, N.~Arkani-Hamed, S.~Dimopoulos and G.~R.~Dvali,
%``New dimensions at a millimeter to a Fermi and superstrings at a TeV,''
Phys.\ Lett.\ B {\bf 436} (1998) 257
[hep-ph/9804398].
%%CITATION = HEP-PH 9804398;%%
%
\bibitem{ads1}
I.~Antoniadis, E.~Dudas and A.~Sagnotti,
%``Supersymmetry breaking, open strings and M-theory,''
Nucl.\ Phys.\ B {\bf 544} (1999) 469
[hep-th/9807011].
%%CITATION = HEP-TH 9807011;%%
%
\bibitem{adds}
I.~Antoniadis, G.~D'Appollonio, E.~Dudas and A.~Sagnotti,
%``Partial breaking of supersymmetry, open strings and M-theory,''
Nucl.\ Phys.\ B {\bf 553} (1999) 133
[hep-th/9812118];
%%CITATION = HEP-TH 9812118;%%
%``Open descendants of Z(2) x Z(2) freely-acting orbifolds,''
Nucl.\ Phys.\ B {\bf 565} (2000) 123
[hep-th/9907184].
%%CITATION = HEP-TH 9907184;%%
%
\bibitem{cotrone}
A.~L.~Cotrone,
%``A Z(2) x Z(2) orientifold with spontaneously broken supersymmetry,''
Mod.\ Phys.\ Lett.\ A {\bf 14} (1999) 2487
[hep-th/9909116].
%%CITATION = HEP-TH 9909116;%%
%
%\cite{pom}
\bibitem{pom}
A.~Pomarol and M.~Quiros,
%``The standard model from extra dimensions,''
Phys.\ Lett.\ B {\bf 438} (1998) 255
[hep-ph/9806263];\\
%%CITATION = HEP-PH 9806263;%%
I.~Antoniadis, S.~Dimopoulos, A.~Pomarol and M.~Quiros,
%``Soft masses in theories with supersymmetry breaking by  TeV-compactification,''
Nucl.\ Phys.\ B {\bf 544} (1999) 503
[hep-ph/9810410];\\
%%CITATION = HEP-PH 9810410;%%
A.~Delgado, A.~Pomarol and M.~Quiros,
%``Supersymmetry and electroweak breaking from extra dimensions at the  TeV-scale,''
Phys.\ Rev.\ D {\bf 60} (1999) 095008
[hep-ph/9812489].
%%CITATION = HEP-PH 9812489;%%
%
\bibitem{bhn}
R.~Barbieri, L.~J.~Hall and Y.~Nomura,
%``A constrained standard model from a compact extra dimension,''
Phys.\ Rev.\ D {\bf 63} (2001) 105007
[hep-ph/0011311].
%%CITATION = HEP-PH 0011311;%%
%
\bibitem{abpss}
C.~Angelantonj, M.~Bianchi, G.~Pradisi, A.~Sagnotti and Y.~S.~Stanev,
%``Chiral asymmetry in four-dimensional open- string vacua,''
Phys.\ Lett.\ B {\bf 385} (1996) 96
[hep-th/9606169].
%%CITATION = HEP-TH 9606169;%%
%
\bibitem{afiv}
G.~Aldazabal, A.~Font, L.~E.~Ibanez and G.~Violero,
%``D = 4, N = 1, type IIB orientifolds,''
Nucl.\ Phys.\ B {\bf 536} (1998) 29
[hep-th/9804026].
%CITATION = HEP-TH 9804026;%%
%
\bibitem{antoniadis}
I.~Antoniadis, K.~Benakli and A.~Laugier,
%``D-brane models with non-linear supersimmetry''
[hep-th/0111209].
%%CITATION = HEP-TH 0111209;%%
%
%\cite{Scrucca:2002is}
\bibitem{paper}
C.~A.~Scrucca, M.~Serone and M.~Trapletti,
%``Open string models with Scherk-Schwarz SUSY breaking and localized  anomalies,''
Nucl.\ Phys.\ B {\bf 635} (2002) 33
[arXiv:hep-th/0203190].
%%CITATION = HEP-TH 0203190;%%
\end{thebibliography}
\end{document}